\documentclass[final,twocolumn,10pt,titlepage,prb]{revtex4}
\usepackage{amsfonts}
\usepackage{amsmath}
\usepackage{amssymb}
\usepackage{graphicx}
\providecommand{\U}[1]{\protect\rule{.1in}{.1in}}
\begin{document}
\preprint{ }

\title[Short title for running header]{Magnetic properties of substitutional Mn in (110) GaAs surface and subsurface 
layers.}
\author{T. O. Strandberg}
\affiliation{Kalmar University, 391 82 Kalmar, Sweden}
\author{C. M. Canali}
\affiliation{Kalmar University, 391 82 Kalmar, Sweden}
\author{A. H. MacDonald}
\affiliation{Department of Physics, University of Texas at Austin, Austin, Texas 78712, USA}
\pacs{PACS number}

\begin{abstract}
Motivated by recent STM experiments, we present a theoretical study of the
electronic and magnetic properties of the Mn-induced acceptor level obtained
by substituting a single Ga atom in the (110) surface layer of GaAs or in one of the
atoms layers below the surface. We employ a kinetic-exchange
tight-binding model in which the relaxation of the (110) surface is taken
into account. The acceptor wave function is strongly anisotropic in space and
its detailed features depend on the depth of the sublayer in which the Mn atom is
located.  The local-density-of-states (LDOS) on the (110) surface associated with the 
acceptor level is more sensitive to the direction of the Mn
magnetic moment when the Mn atom is located 
further below the surface. 
We show that the total magnetic anisotropy energy of the
system is due almost entirely to the dependence of the acceptor level energy on Mn spin orientation,
and that this quantity is strongly dependent on the depth of the Mn atom.

\end{abstract}
\volumeyear{2009}
\volumenumber{number}
\issuenumber{number}
\eid{identifier}
\date[Date text]{date}
\received[Received text]{date}

\revised[Revised text]{date}

\accepted[Accepted text]{date}

\published[Published text]{date}

\startpage{101}
\endpage{102}
\maketitle

\section{Introduction}

Cross sectional scanning tunneling microscopy (STM) is a 
sophisticated and extremely valuable nanoscience experimental tool. 
It enables, in particular, the
manipulation and characterization of individual impurities in semiconductors
and metals with unprecedented spatial resolution and electronic
sensitivity. In the last few years this technique has been used to study the
electronic and magnetic properties of individual\cite{yakunin_prl04}
Mn atoms and Mn atom pairs\cite{yakunin_prl05} in GaAs. Interest in these studies stems
in part from the notion that investigating Mn dopants in GaAs at the atomic
scale could lead to a better understanding of and better control over magnetism in diluted
magnetic semiconductors (DMSs). Recently STM methods were used to substitute
individual Ga atoms with Mn atoms in the first layer of a GaAs(110)
surface.\cite{kitchen_jsc05,yazdani_nat06} STM allowed experimentalists to
visualize the electronic properties of the acceptor wavefunction bound to the Mn
position, and to probe the valence-band-hole mediated exchange interactions between two Mn
atoms.\cite{yazdani_nat06} Apart from its possible relevance in probing 
the basic physics of DMSs, this experiment is a remarkable example of how STM
techniques can now be employed to engineer novel nanomagnets with specifically
designed quantum mechanical properties. 

In Refs.~[\onlinecite{kitchen_jsc05, yazdani_nat06}] it was
shown that both the Mn-induced (acceptor) in-gap state and the exchange
interaction between Mn pairs are strongly anisotropic with respect to
crystallographic orientation. A theoretical analysis\cite{yazdani_nat06} based
on a tight-binding model of Mn atoms in \textit{bulk}
GaAs\cite{tangflatte_prl04,tangflatte_prb05} reproduces some of
these features qualitatively.  Nevertheless, 
it seems clear that the proximity of the Mn atoms to the surface 
must have a significant impact on their properties. In fact, a more recent
study\cite{koenraad_prb08}, in which individual Mn atoms were carefully
positioned on the GaAs(110) surface and layer-by-layer on the first few
layers below, showed that the reduced symmetry at the surface strongly
modifies the wave function of the impurity.  The acceptor wavefunction properties depend on the
precise substitutional depth at which the impurity is located. 
This experimental conclusion is supported by a recent theoretical study.\cite{jancu_prl08}

In this paper we present a theoretical study of individual Mn dopants
substituting for Ga atoms in a GaAs(110) surface, or in one of 
atomic layers below the surface. Our aim is to provide a systematic
analysis of how the electronic and magnetic properties of the acceptor
wavefunction are modified by the presence of the surface as the Mn impurity is
inserted into successively deeper layers, and compare with the limiting case
of a Mn in bulk GaAs. In contrast to previous studies\cite{jancu_prl08}, we
focus on the spin-orbit induced dependence of the acceptor wave function on
the direction of the Mn magnetic moment orientation.
We compare our surface-influenced results with the
results of Ref.~\onlinecite{tangflatte_prb05}, in which the dependence of the
acceptor wavefunction on the magnetic moment direction was studied for a Mn
atom in bulk GaAs. 

Our analysis is based on a microscopic tight-binding
model, which 
accounts for the crucial relaxation of the GaAs(110) surface layer and for  
spin-orbit interactions in the valence band which play an essential\cite{jungw_rmp06} role 
in (Ga,Mn)As magnetism.
We do not account explicitly for the Mn $d$-orbitals, but account for 
$d-p$ hybridization instead by adding 
an effective exchange interaction between the Mn moment and valence band 
orbitals on 
nearest-neighbor As sites.
We also include other interaction terms to account for the Coulomb  
repulsion of electrons by the Mn ion. 

Our calculations show
that the acceptor wavefunction is in general strongly anisotropic in space;
the detailed anisotropy features depend very strongly on the sublayer in which the
impurity is located, in agreement with experiment\cite{koenraad_prb08} and with
previous calculations.\cite{jancu_prl08} For a reasonable choice of the
parameters of our model, we find that for a Mn located in the topmost layer
or in the first subsurface layer, the acceptor state has a large
binding energy and a strongly localized wavefunction with a very weak
dependence on the Mn magnetic moment direction. As the impurity is inserted
deeper beneath the surface, the acceptor wavefunction becomes progressively
more delocalized and its dependence on the Mn moment orientation increases
significantly. In particular, we find that the LDOS feature on the
GaAs(110) surface due to the acceptor level, the quantity which is probed 
most directly by STM, is noticeably different between the cases of a
magnetic moment pointing along the easy and hard magnetic directions
when the impurity is located a few
monolayers below the surface.  This prediction could be tested in STM experiments 
in which the direction of the magnetic moment is manipulated with an external magnetic
field. We also show that total magnetic anisotropy is related in a simple way to the
magnetic anisotropy of the acceptor state and that 
the magnetic anisotropy landscape depends in a non trivial way on the sublayer
in which the impurity is located.  

Our paper is organized as follows. In Sec.~\ref{theory} we give an explicit 
description of the model we use, explaining its tight-binding Hamiltonian and elaborating on other 
details necessary to understand our findings. In Sec.~\ref{results}
we present the results of the model, starting in Sec.~\ref{results_bulk} with
the case of a single Mn substituted for Ga in bulk GaAs. We then proceed in
Sec.~\ref{results_surface} to discuss a substitutional Mn in a $\left(  110\right)  $
surface layer, highlighting emerging features and relating these to experiment.
Sec.~\ref{surface_to_bulk} is devoted to the study of the transition from
surface to bulk, in which the Mn is placed in successively deeper sublayers
below the surface. Finally, we present our conclusions and discuss the implications
of our results in Sec.~\ref{conclusions}. 
The main text of this paper provides a detailed 
description of the Mn impurity at different depths below the $\langle110\rangle$ surface
layer of the host semi-conductor. Some readers 
may wish to begin by reading the summary and conclusion section, which gives a brief description of
our main results, before exploring the main body of the paper. 

\section{Theory}

\label{theory}In a $\left(  \text{Ga,Mn}\right)  $As III-V dilute magnetic
semiconductor (DMS), the most energetically stable position for the Mn is a Ga
atom site. This property can be understood as following from the atomic
electronic structure of Mn, $[Ar]3d^{5}4s^{2}$, with the $4s^{2}$ electrons
allowing for the formation of crystal bonds similar to those between the Ga
($[Ar]3d^{10}4s^{2}p^{1}$) and the As ($[Ar]3d^{10}4s^{2}p^{3}$) atoms of the
host crystal. Because the Mn is missing the 4$p$ valence electron of Ga, it
acts as an acceptor. The Mn ion repels electrons and attracts a weakly bound
hole, forming a neutral state.\cite{kitchen_jsc05,yakunin_pae05} Mn both
generates the local magnetic moments (via the $3d^{5}$ half-filled d-shell)
and acts as a supplier of potentially itinerant holes that can mediate their
coupling. Zener's kinetic-exchange \cite{zener_pr51,dietl_sci00,jungw_rmp06}
or indirect-exchange interaction applies to systems like (Ga,Mn)As in which
local moments formed by the magnetic impurities are coupled via itinerant $s$-
or $p$-band carriers.

Our study is based on tight-binding model with a Hamiltonian
\begin{equation}
H=H_{\mathrm{band}}+H_{\mathrm{exc}}+H_{\mathrm{SO}}+H_{\mathrm{coul}} \;.
\label{hamiltonian}
\end{equation}
that includes a kinetic exchange interaction between the local-moment and the
band electrons. The kinetic-exchange model is appropriate\cite{jungw_rmp06}
when the Mn-$d$ to As-$p$ hopping amplitudes (see below) are smaller than the
energetic separation between the $d$-orbitals and the top of the valence band.
The band term in Eq.~\ref{hamiltonian} is given in terms of the Slater-Koster
parameters\cite{slaterkoster_pr54,papac_jpcm03} for bulk
GaAs.\cite{chadi_prb77}
\begin{equation}
H_{\mathrm{band}}=\sum_{ij}\sum_{\mu\mu^{\prime}}\sum_{\sigma}t_{\mu
\mu^{\prime}}^{ij}a_{i\mu\sigma}^{\dag}a^{\phantom {\dag}}_{j\mu^{\prime
}\sigma}\;.
\end{equation}
Here, $i$ and $j$ are atomic indices, and $\mu$ and $\sigma$ are orbital and
spin indices, respectively. The $t_{\mu\mu^{\prime}}^{ij}$ are the
Slater-Koster parameters that do not depend on spin. The only non-zero
parameters are the on-site energies ($i=j,$ $\mu=\mu^{\prime}$) and the
nearest neighbor hopping matrix elements for the $s,$ $p_{x}$, $p_{y}$ and
$p_{z}$ orbitals.

Our model contains $s$ and $p$ electrons only, the $d$-electrons of the Mn
enter the model in the form of a Mn local moment with spin $S=5/2$ spin which
we treat classically in this paper. This local magnetic moment is formed by
the five $3d$-electrons of Mn that in the tetrahedral host results in bonding
and antibonding $sp$-$d$ states in the form\cite{vogl_app05} of a triplet of
$t_{2g}$-symmetry ($3d_{xy}$-$,3d_{zx}$- and 3$d_{yz}$-like) and
an occupied doublet of $e_{g}$-symmetry ($3d_{x^{2}-y^{2}}$- and $3d_{z^{2}}
$-like). The doublet couples only weakly to the host, and is split from the
triplet by the tetrahedral crystal field. The triplet hybridizes with the
connecting $sp$-orbitals and the weakly bound hole occupies one of the three
$sp$-$d$ antibonding states at the top of the valence band, predominantly of
As 4$p$-character.\cite{jungw_rmp06} The hybridization of the occupied Mn
3$d$-electrons with the nearest neighbor As 4$p$-electrons, cause the
$p$-states at the top of the valence band with spin parallel to the Mn spin to
move up in energy relative those that are antiparallel, which hybridize with
high-energy unoccupied $d$-orbitals. The direct exchange interaction between
holes at the top of the valence band and the Mn $d$-electrons is weak, such
that $p$-$d$ hybridization dominates which results in an antiferromagnetic
coupling.\cite{bhattacharjee_pbc83,okabayashi_prb98} This physics is captured
in the $H_{\mathrm{exc}}$ term of (\ref{hamiltonian}), which induces an
exchange field on the nearest-neighbor As $p$-electrons,
\begin{equation}
H_{\mathrm{exc}}=J_{pd}\sum_{m}\sum_{n[m]}\vec{S}_{n}\cdot\hat{\Omega}_{m}\;,
\label{exchange}
\end{equation}
where $J_{pd}=1.5$ eV is the approximate value of the exchange coupling
constant inferred from Refs.~[\onlinecite{timmacd_prb05}]
and~[\onlinecite{ohno_sci98}], and
\begin{equation}
\vec{S}_{n}=\tfrac{1}{2}~\sum_{\pi\sigma\sigma^{\prime}}a_{n\pi\sigma}^{\dag
}\vec{\tau}_{\sigma\sigma^{\prime}}a_{n\pi\sigma^{\prime}}^{\phantom {\dag}}
\;, \label{pspin}
\end{equation}
where $\vec{\tau}$ is the Pauli matrix vector. The first sum over $m$ in
(\ref{exchange}) runs over all Mn, and the second sum runs over all As that
are nearest neighbors to Mn atom $m,$ denoted by $n[m]$. Because the exchange
field in this model influences only $p$-electrons, the sum $\pi$ in
(\ref{pspin}) only runs over the three $p$-orbitals. The valence band
electronic structure depends on the classical Mn spin orientation $\hat
{\Omega}_{m}$ (which we parametrized by polar $\theta$ and azimuthal $\phi$
angles) through the scalar product $\vec{S}_{n}\cdot\hat{\Omega}_{m}$.

We approximate the spin-orbit coupling
Hamiltonian $H_{\mathrm{SO\text{ }}}$ by a local atomic one-body operator, in
which the spin quantization axis is defined by $\hat{\Omega}_{m}$:
\begin{equation}
H_{\mathrm{SO}}=\sum_{i}^{{}}\sum_{\mu,\mu^{\prime},\sigma,\sigma^{\prime}
}\lambda_{i}\langle\mu,\sigma|\vec{L}\cdot\vec{S}|\mu^{\prime},\sigma^{\prime
}\rangle a_{i\mu\sigma}^{\dag}a_{i\mu^{\prime}\sigma^{\prime}}
^{\phantom {\dag}}\;,
\end{equation}
where $i$ is an atomic index and $\lambda_{i}$ denotes the renormalized
spin-orbit splitting\cite{chadi_prb77} for which we use the values
$\lambda_{i\in\{Ga\}}=58$ meV, $\lambda_{i\in\{As\}}=140$ meV, and
$\lambda_{i\in\{Mn\}}=\lambda_{i\in\{Ga\}}/2$. The spin-orbit term causes the
total energy of the system, as obtained by summing the energies of all
occupied eigenstates, to depend on the magnetization direction parametrized
by $\hat{\Omega}_{m}$. Our procedure allows us to calculate the magnetic
anisotropy landscape on the unit sphere and extract the magnetic anisotropy
energy as $E_{\mathrm{anis}}=\max E\left(  \theta,\phi\right)  -\min E\left(
\theta,\phi\right)  $. The polar coordinate system is defined such that
$\theta=0$ corresponds to the [001] direction, and with $\theta=\pi/2,\phi=0$
and $\phi=\pi/2$ correspond to the [100] and the [010] directions respectively.

The presence of a negatively charged Mn ion attracts holes and repels
electrons. We represent the spin-independent part of the effective potential
due to Mn substitution by a long-range repulsive Coulomb part and a Mn central
cell correction term,
\begin{equation}
H_{\mathrm{coul}}=\frac{e^{2}}{4\pi\varepsilon_{0}\varepsilon_{r}}\sum_{m}
\sum_{i\mu\sigma}\frac{a_{i\mu\sigma}^{\dag}a_{i\mu\sigma}^{\phantom {\dag}}
}{|\vec{r}_{i}\mathbf{-}\vec{R}_{m}|}+V_{\mathrm{corr}}\;. \label{coulomb}
\end{equation}
The first term in (\ref{coulomb}) represents the long-range part, which is
reduced by the host material dielectric constant $\varepsilon_{r}=12.$ To
account crudely for weaker dielectric screening at the surface, the dielectric
constant for a Mn on the surface is reduced to $\varepsilon_{r}=6$ for the
affected surface atoms. The correction term consists of on- and off-site
parts, $V_{\mathrm{corr}}=V_{\mathrm{on}}+V_{\mathrm{off}}$ which influence
the Mn ion and its nearest neighbors respectively. The on-site Coulomb
correction is estimated to $1.0$ eV from the ionization energy of Mn. The
off-site Coulomb correction affects all the nearest-neighbor As surrounding
the Mn ion and together with $H_{\mathrm{exc}}$ (\ref{exchange}) reflects
primarily $p$-$d$ hybridization physics. It is one of the most important
parameters of the model and its value is set by tuning the position of the
Mn-induced acceptor level in the bulk to the experimentally observed
position\cite{schairer_prb74,lee_ssc64,chapman_prl67,linnarsson_prb97} at 113
meV above the first valence band level. The value thus obtained is
$V_{\mathrm{off}}=2.4$ eV. In this picture, long-range Coulomb, exchange and
correction interactions all play an important role in determining the
character of Mn acceptor levels.

We model the electronic structure of GaAs with a single substitutional Mn by
performing a super-cell type calculation with a cubic cluster of 3200 atoms
and periodic boundary conditions in either 2 or 3 dimensions, depending on
whether we are studying the $\left(  110\right)  $ surface or a bulk-like
system. The $\left(  110\right)  $ surface of GaAs is simplified from both
theoretical and experimental points of view, by the absence of large surface
reconstruction. Relaxation of surface layer positions must nevertheless be
included since it removes dangling-bond states that would otherwise obscure
the band-gap. We follow the procedure outlined in
Refs.~[\onlinecite{chadi_prl78, chadi_prb79}], in which atomic shifts as deep
as the 2nd sublayer are taken into account. \begin{figure}[ptb]
\resizebox{6cm}{!}{\includegraphics{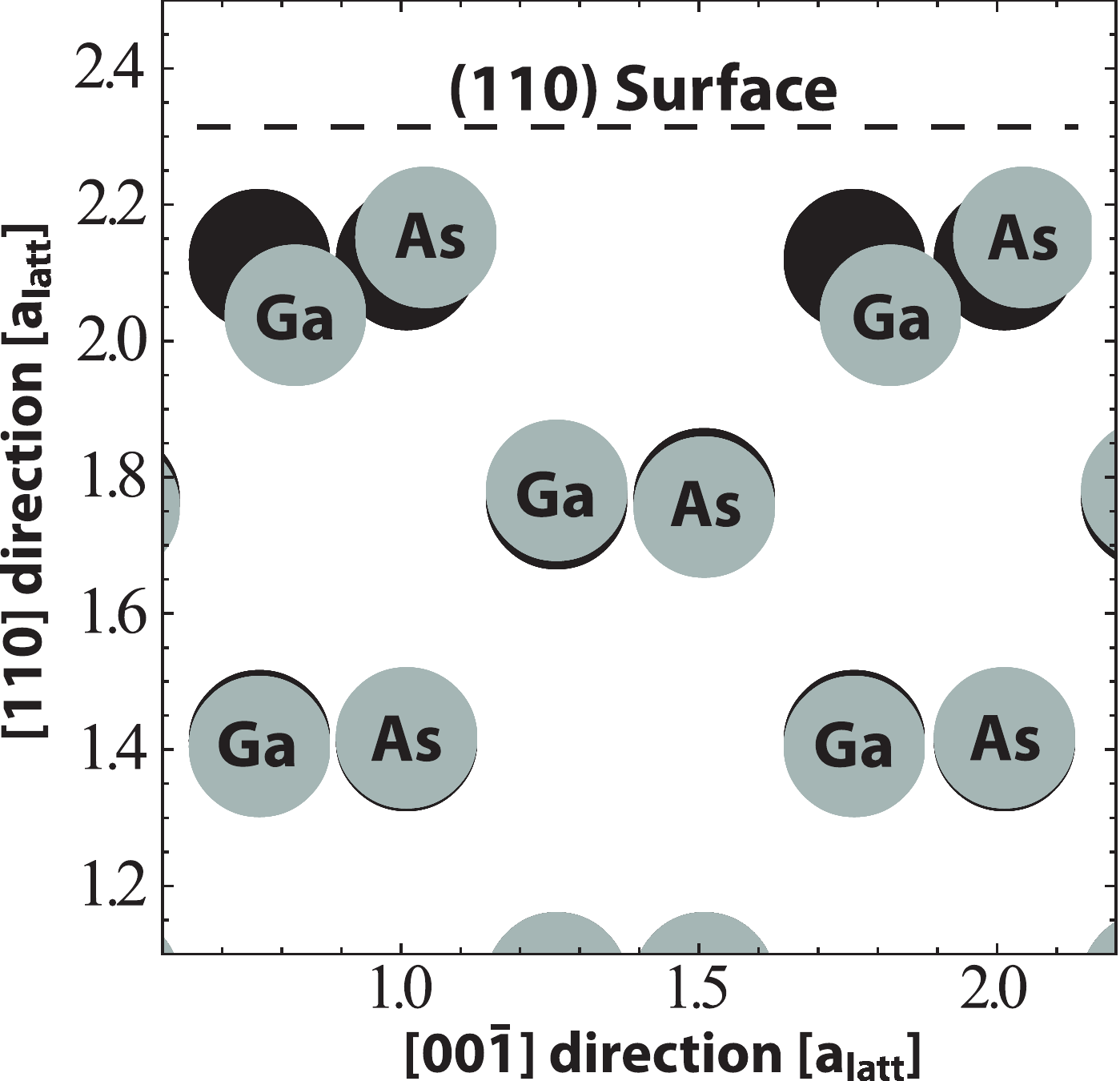}}\caption{The
relaxed (110) surface. Gray filled circles signify relaxed
positions and the black filled circles the unrelaxed positions. This
illustration shows a side view of the (110) surface with distances in units of
the GaAs lattice constant $a_{\mathrm{latt}}=0.565$ nm. }
\label{relaxfig}
\end{figure}\begin{figure}[ptb]
\resizebox{6cm}{!}{\includegraphics{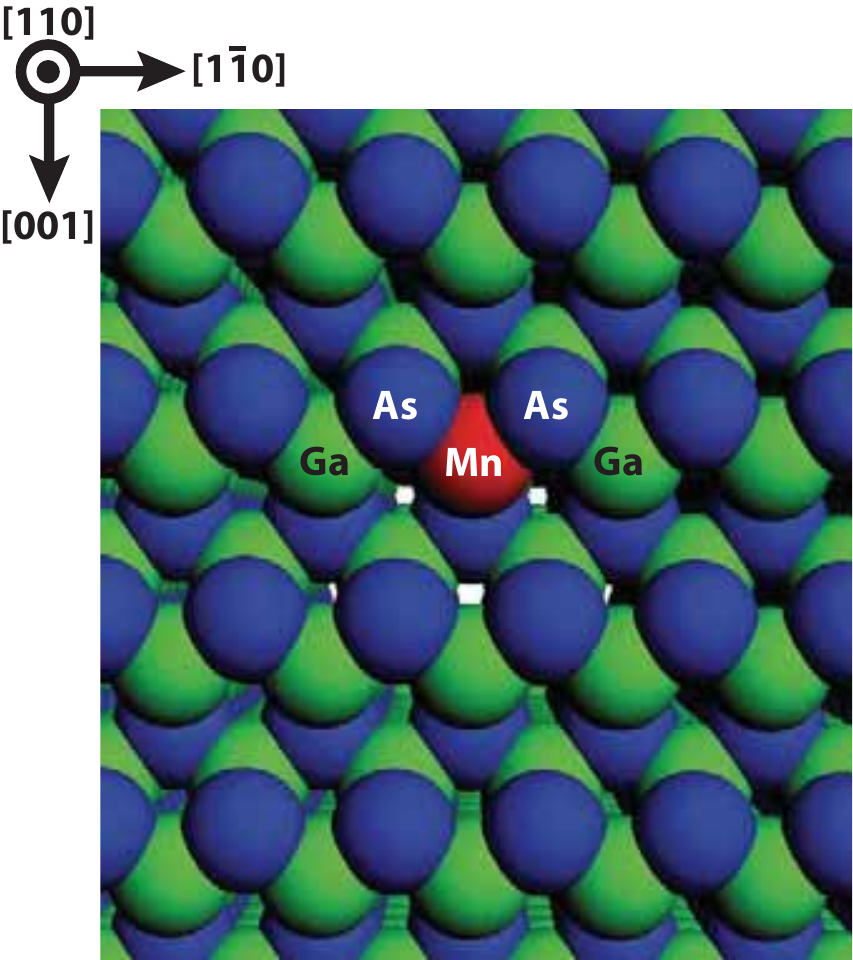}}\caption{(Color online) The (110)
surface. The green spheres represent Ga, blue spheres the As and red sphere
indicates a Mn that has replaced a Ga.}
\label{clusterfig}
\end{figure}The resulting $\left(  110\right)  $ surface
is summarized in Fig.~\ref{relaxfig} and the top layer and first subsurface
layer are depicted in Fig.~\ref{clusterfig}. The most salient feature is an
up-shift of the surface As, accompanied by a down-shift of the surface Ga.
Both species are shifted along the surface, such that the effective
bond-angles change, which affects the direction cosines entering the
Slater-Koster parametrization.\cite{slaterkoster_pr54} The tight-binding
parameters scale like $d_{0}^{2}/d^{2}$, where $d_{0}$ is the original
distance and $d$ is the distance after the relaxation. The rescaling parameter
is small and is at most $d_{0}^{2}/d^{2}\approx1.06.$

By projecting the eigenvector of the acceptor level onto surface sites
obtained after diagonalizing the Hamiltonian $\left(  \ref{hamiltonian}
\right)  $, we can study the surface LDOS, which is intimately related to the
topographic STM images.\cite{koenraad_prb08,yazdani_nat06} In our cluster
approach we sum the absolute square of the coefficients in the eigenvector
corresponding to the LDOS on a given atom $j$,
\begin{align}
\langle\psi_{n}|P_{j}|\psi_{n}\rangle &  =\sum_{i\mu\sigma,i^{\prime}
\mu^{\prime}\sigma^{\prime}}c_{i^{\prime}\mu^{\prime}\sigma^{\prime}}^{n\ast
}c_{i\mu\sigma}^{n}\langle i^{\prime}\mu^{\prime}\sigma^{\prime}|P_{j}
|i\mu\sigma\rangle\label{projatom}\\
&  =\sum_{\mu\sigma}|c_{j\mu\sigma}^{n}|^{2}\;,
\end{align}
where $n$ is an eigenvalue index, $i$ and $j$ are atomic indices, and $\mu$
and $\sigma$ denote orbital and spin respectively. In (\ref{projatom})
$P_{j}=\sum_{\mu\sigma}|j\mu\sigma\rangle\langle j\mu\sigma|$ projects out the
LDOS of atom $j.$ Similarly, we can define operators that project out the
orbital and spin character for a given eigenlevel $n$:
\begin{align}
P_{\mu}  &  =\sum_{i\sigma}|i\mu\sigma\rangle\langle i\mu\sigma
|\;,\label{orbproj}\\
P_{\sigma}  &  =\sum_{i\mu}|i\mu\sigma\rangle\langle i\mu\sigma|\;.
\label{spinproj}
\end{align}
The procedure we follow in generating LDOS plots is similar to the approach
used by Tang~\textit{et al.} in Ref.~\onlinecite{tangflatte_prl04}. We place
Gaussians on the atomic positions with a magnitude equal to the LDOS of that
atom, and a full-width at half-maximum equal to half the nearest neighbor
spacing. (This procedure mimics the finite spatial resolution of an STM tip).
In all the LDOS plots the normalization is such that the sum of LDOS over all
atoms in the cluster is unity for a single eigenlevel. When the STM is
operating in constant current mode (see for example
Ref.~\onlinecite{wiesend_98}), the tunneling current is maintained at a fixed
value by varying the tip-surface distance. The exponential decay of the
surface wavefunction causes the distance recorded to depend approximately
logarithmically on the LDOS at the surface. We therefore employ a logarithmic
color-scale in our images of the LDOS.

When imaging states in a semi-conductor band gap at a low bias, it is
necessary to move the tip very close to the substrate surface. This means that
interactions between the tip and the sample can cause a change in the surface
wavefunction.
In addition, many-body effects beyond those captured by the mean-field
description of electronic states outlined above can be important in some
cases. In particular, as detailed below, we find that acceptor levels for Mn
very close to the surface lie deep in the gap. STM experiments provide a
partial profile of the spatial distribution of added or removed electrons.
When the acceptor is deep in the gap, addition or removal will cause a big
change in the potential seen by other electrons. These various many-body
effects are partly captured by our phenomenological model, but we cannot
expect to find exact correspondence with the experimental images.

\section{Results}

\label{results}

\subsection{Single Mn in bulk GaAs}

\label{results_bulk}

Our starting point is a single Mn in bulk, as represented in our model by
placing it at the center of a cubic GaAs cluster of 3200 atoms and enforcing
periodic boundary conditions. The dimensions of the cluster (in the
crystalline directions $[110]\times\lbrack1\bar{1}0]\times\lbrack001]$) used
in all calculations are $38.0\times38.0\times42.4$\AA $^{3},$ or in terms of
atomic layers $20\times20\times32$. Creating the supercell in the form of a
cubic cluster defined by the limiting planes $(110)$, $(1\bar{1}0)$ and
$(001)$, enables us to apply periodic boundary conditions in two directions
and study the (110) surface. The generated cluster can also be given periodic
boundary conditions in all three directions to create a 'bulk' system, which
should simply be regarded as the fully periodic counterpart of the surface
system. \begin{figure}[ptb]
\resizebox{7.5cm}{!}{\includegraphics{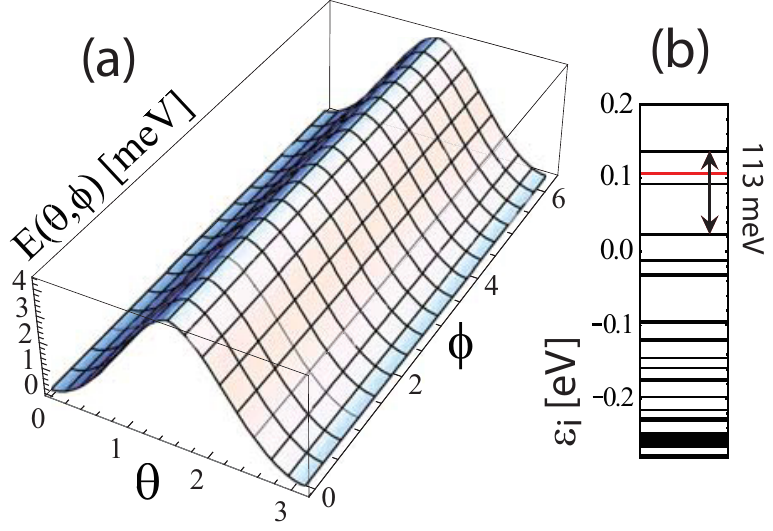}}\caption{(Color online) In (a) the
magnetic anisotropy energy of a single Mn in a 3200 atom cluster with periodic
boundary conditions. Bistable easy directions parallel to the [001] axis are
separated by a single barrier of magnitude 4.35 meV in the (001) plane.
(b) shows the ground-state energy level spectrum in the easy direction,
where the off-site Coulomb correction term been used to tune the
acceptor level at 113 meV above the first valence band level. }
\label{bulkmae}
\end{figure}

The magnetic anisotropy energy for the fully periodic system as a function of
the magnetization direction $E(\theta,\phi)$, is shown in Fig.~\ref{bulkmae}.
We find bistable minima and an easy axis parallel to the $[$001$]$ direction,
separated by a single barrier equal to $E_{\mathrm{anis}}=4.35$ meV. The
anisotropy is very sensitive to the cubic symmetry and the particular easy
direction $[$001$]$ can be seen as a consequence of the supercell symmetry.
The closest distance between Mn in adjacent supercells is 38 \AA \ along the
[110] and [1\={1}0] directions, along which the most prominent hopping path
occurs via closely spaced As and Ga. The distance along the 4 other equivalent
symmetry directions is much longer, resulting in the observed anisotropy.
Scaling down the size of the supercell (and thereby increasing the effective Mn
doping) to 1200 atoms yields an anisotropy of 8.8 meV, and the value increases
further as the supercell size is decreased.
We conclude from this calculation that the magnetic anisotropy of a single Mn
in the bulk of an infinite crystal is much smaller than $\sim4$ meV, and that,
as far as magnetic anisotropy is concerned, the dilute \emph{isolated
impurity} limit is achieved only at Mn atom fractions $x$ much smaller than
$10^{-3}$, and much smaller than what can be represented in this or any other
supercell calculation.

\begin{figure}[ptb]
\resizebox{8.5cm}{!}{\includegraphics{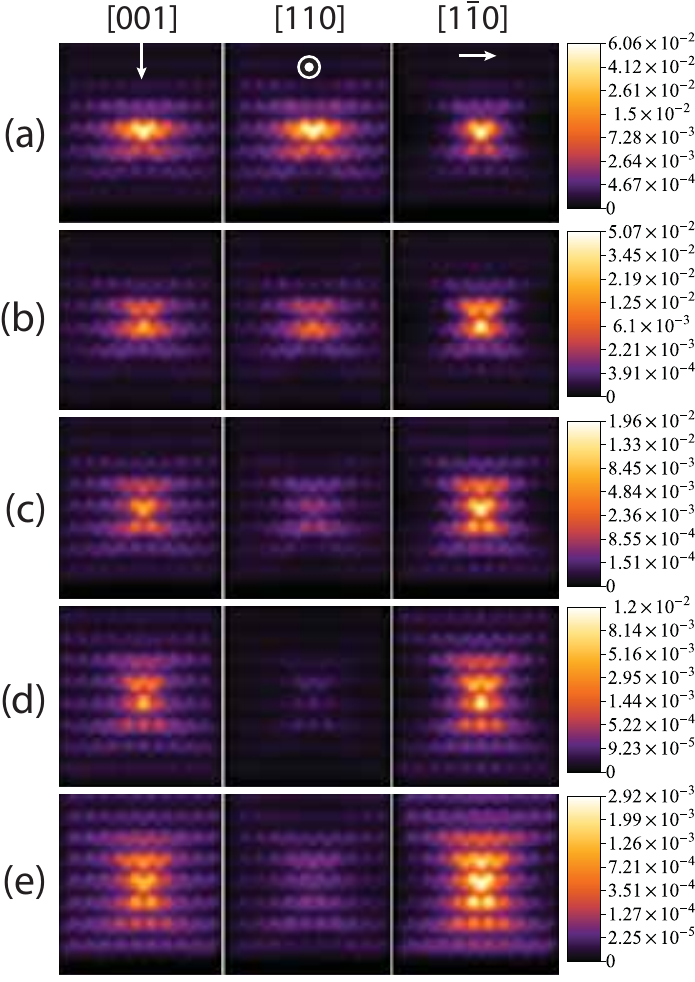}}
\caption{(Color online) The bulk acceptor level LDOS of a single Mn in a cluster with
periodic boundary conditions in all directions. The figures show (110) plane
cuts of the LDOS containing the Mn in (a), and moving from one to four layers
away from it in (b)-(e). The direction of the Mn spin is indicated by white
arrows at the top of the figure. The left column shows the LDOS when the Mn
spin is in the easy [001] direction, and the center and right column when the
Mn spin is in two hard directions, $[$110$]$ and $[$1\={1}0$]$. The successive
cuts reveal a decreasing magnitude of the LDOS with distance from the impurity
and a strong dependence on the Mn spin orientation. Note the alternating
behavior on odd and even layers away from the Mn plane. When the Mn spin is in
the easy direction, the acceptor wavefunction spreads symmetrically in the
$[$110$]$ and $[$1\={1}0$]$ directions. Comparing the case when the spin is in
a direction parallel to the (110) plane normal (center column), and when the
spin is perpendicular to the (110) plane normal (right column), it can be seen
that the wavefunction is highly anisotropic and extends in along a symmetry
direction perpendicular to the spin direction.}
\label{bulkldos}
\end{figure}
Fig.~\ref{bulkldos} shows plots of the LDOS of the acceptor level
in the easy and hard directions. Similar results have been obtained previously
by Tang \textit{et al.}~
\cite{koenraad_prb08,yazdani_nat06,tangflatte_prl04,tangflatte_prb05} using a
similar tight-binding model. When the Mn spin is pointing in the easy
direction [001], the wavefunction spreads out symmetrically along $[$110$]$
and $[$1\={1}0$]$. The spread along $[$110$]$ is shown is shown in
Fig.~\ref{bulkldos} (a)-(e) in the left column, in which cuts of successive
(110) planes up to 4 layers away from the Mn plane are displayed. These images
agree qualitatively with the results presented in
Ref.~\onlinecite{koenraad_prb08}, in which statistical methods and comparison
with theory enabled the identification of the location of a particular single
Mn down to the fourth sublayer. The center and the right column in
Fig.~\ref{bulkldos} show the LDOS in successive (110) planes when the Mn spin
is oriented in the $[$110$]$ and $[$1\={1}0$]$ hard directions respectively.
The images show that the acceptor wavefunction exhibits a definite preference
to spread out along the symmetry direction perpendicular to the Mn spin. In
the center column of Fig.~\ref{bulkldos} it can be seen that the spread along
the Mn spin direction [110] is very weak. By contrast, when the Mn spin is
pointing in the $[$1\={1}0$]$ direction, perpendicular to the (110) plane
normal, the rightmost column of Fig.~\ref{bulkldos} reveals much higher values
of the LDOS.

The cut of the Mn plane in the easy direction (Fig.~\ref{bulkldos} (a)) has 6\%
of the spectral weight of the acceptor wavefunction on the Mn atom, a total of
15.4\% on the 4 nearest neighbors As and the rest is spread out in the
lattice. These values correspond to a more spread out LDOS than the one
obtained by Tang~\textit{et al.} in Ref.~\onlinecite{tangflatte_prl04}, who
find 10\% on the Mn and 20\% on the 4 surrounding As.
In both cases model parameters were adjusted to give the correct energetic
position for the bulk Mn acceptor level. The difference in wavefunctions
demonstrates that satisfying this criterion does not guarantee that the
character of the acceptor level is correctly captured. In particular,
accounting for the contribution to binding from longer ranged Coulomb
interactions lead to more extended wavefunctions at a given acceptor energy.
The observed correlation between spatial anisotropy and Mn spin direction
agrees qualitatively with previous calculations\cite{tangflatte_prb05}, in
which a decrease of 90\% in maximum spectral weight at the center of the (110)
plane images is seen four layers away from the Mn (corresponding to
Fig.~\ref{bulkldos} (e)), when the Mn spin changes from $[$1\={1}0$]$ to [110],
and a decrease of 15\% when the Mn spin changes from $[$1\={1}0$]$ to [001].
Comparing this with our results we find similar anisotropies. To begin with
consider the anisotropy of the wavefunction at three layers away from the Mn
(Fig.~\ref{bulkldos} (d)). Here the maximum LDOS in the (110) plane drops by
86\% as the spin changes from $[$1\={1}0$]$ to [110], and by 21\% when the
spin changes from $[$1\={1}0$]$ to [001]. Looking at the next layer
(Fig.~\ref{bulkldos} (e)), we note that this anisotropic effect decreases
slightly, and we find that the maximum LDOS decreases by 74\% when the spin
direction is changed from $[$1\={1}0$]$ to [110], and by 25\% when it is
changed from $[$1\={1}0$]$ to [001].
The source of this difference in the behavior of the LDOS between odd and even
layers away from the Mn, is that the odd layer maximum is on the As and the
even layer maximum is on the Ga in the nearest neighbor hoping path along the
[110]. This effect becomes pronounced three layers away from the Mn and
higher. Although the actual percentages differ somewhat, the same strong
anisotropic behavior of the acceptor wavefunction as in
Ref.~[\onlinecite{tangflatte_prb05}] is observed.
The strong dependence of the LDOS on Mn spin-orientation, combined with the
weak magnetic anisotropy, implies that the observed LDOS should be strongly
sensitive to external magnetic fields.

\begin{figure}[ptb]
\resizebox{7cm}{!}{\includegraphics{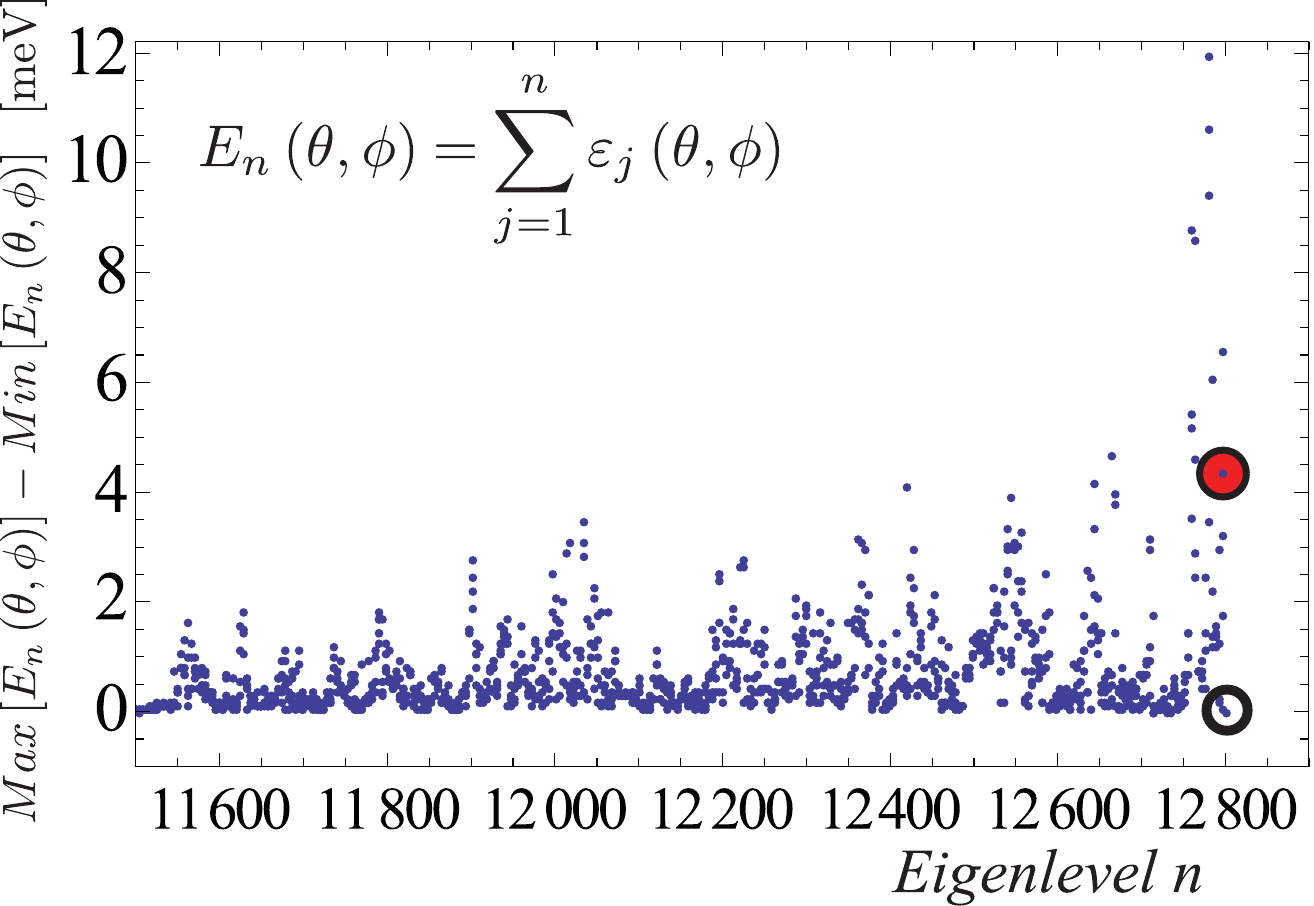}} \caption{The
total magnetic anisotropy energy obtained by summing the dependence of energy
on magnetization orientation for all occupied orbitals as a function of the
index of the Fermi level orbital. This figure shows that all orbitals in the
valence band are dependent on magnetization orientation, causing the
anisotropy to fluctuate substantially when one occupation number is changed.
The systems total anisotropy energy is indicated by a filled circle. We also
see that the anisotropy is nearly zero when all valence band orbitals are
occupied (indicated by empty circle). The anisotropy energy of a Mn acceptor
is therefore just the anisotropy of the top level in the valence band, with a
change of sign.}
\label{bulkcumul}
\end{figure}The anisotropy energy is formed by summing up the energies of all
occupied levels, counting 4 electrons per Ga and As and 3 electrons per Mn.
The spin-orbit induced level shifts vary with magnetization direction on the
unit sphere and give rise to an anisotropic dependence of the total energy on
the magnetization direction.
Fig.~$\ref{bulkcumul}$
shows the anisotropy energy as obtained by successively summing eigenlevel
anisotropy landscapes on the unit sphere, starting from the lowest level. It
reveals fluctuations that persist far into the valence band.
Fig.~$\ref{bulkcumul}$ illustrates the advantage of using a hole rather an
electron picture in analyzing the anisotropy; the anisotropy built up by the
shifts of many occupied levels, is retrieved by the single unoccupied acceptor
level. This picture remains valid as long as the coupling to the conduction
band is not sensitive to magnetization orientation.
Quite generally we find that the anisotropy of the acceptor level $\varepsilon
_{acc}\left(  \theta,\phi\right)  $ and the systems total energy
$E(\theta,\phi)$ are accurately related by
\begin{align}
&  \varepsilon_{acc}\left(  \theta,\phi\right)  -Min[\varepsilon_{acc}\left(
\theta,\phi\right)  ]\nonumber\\
&  =-\left(  E(\theta,\phi)-Min[E(\theta,\phi)]\right)  . \label{accrel}
\end{align}
This relation holds for several Mn in the system, subject to the same
conditions. The single level $\varepsilon_{acc}\left(  \theta,\phi\right)  $
should then be replaced with a sum over all acceptor levels. Because the
number of occupied levels that contribute to 
the total anisotropy varies from
case to case, the hole picture is invariably more useful in trying to
understand trends.

\begin{figure}[ptb]
\resizebox{6.5cm}{!}{\includegraphics{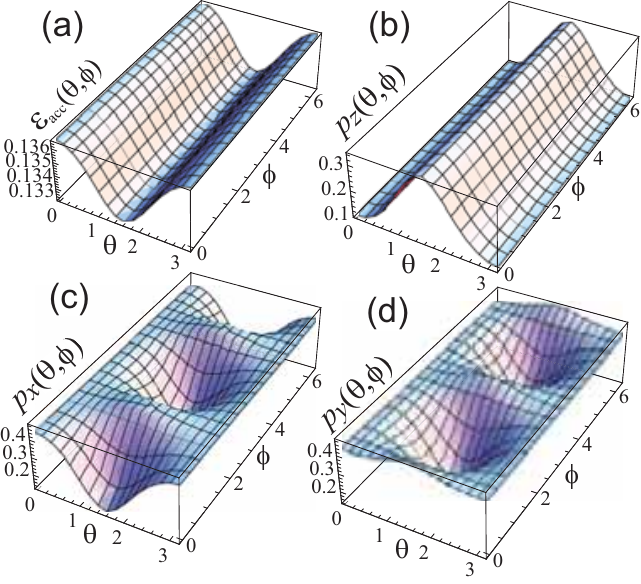}}\caption{(Color online) The
acceptor level and $p$-character variation on the unit sphere. The acceptor
level anisotropy (a) cancels the anisotropy built up by all occupied levels.
This means that the variation of the acceptor level as a function of the
magnetization direction is the negative of the total anisotropy landscape.
Variations in the orbital character (c-d) of the acceptor correlate with an
acceptor wavefunction that spreads mainly in direction perpendicular to the Mn
spin.}
\label{bulkorbitals}
\end{figure}

Fig.~$\ref{bulkcumul}$ shows that the anisotropy energy built up by all
occupied levels (filled circle), is canceled by adding the anisotropy of the
single acceptor level (empty circle). In Fig.~\ref{bulkorbitals} (a) the
acceptor level variation as a function of magnetization direction is shown,
and it can be seen that taking the negative of this reproduces the total
energy variation shown in Fig.~\ref{bulkmae}. The property observed in the
LDOS images (see Fig.~\ref{bulkldos}) that the acceptor wavefunction tends to
spread out in a plane perpendicular to the Mn spin direction is reflected in
the orbital character variation of the acceptor level. Fig.~\ref{bulkorbitals}
(b)-(c) shows the orbital $p$-character of the acceptor level, as obtained by
using the projector (\ref{orbproj}). To illustrate this point, consider the Mn
spin in the direction [001], defined by $\theta=0$. Here, the $p_{z}
$-character of the acceptor wavefunction is lowest and the $p_{x}$ and $p_{y}$
characters are high, consistent with an acceptor wavefunction that is
spreading out in the (001) plane. In the [100] direction, defined by
$\theta=\pi/2$ and $\phi=0$, the $p_{x}$ character is low and the $p_{z}$ and
$p_{y}$ characters are high, as the wavefunction is spreading mainly in the
(100) plane. A similar dip in the $p_{y}$-character is found in the [010]
direction. The sum of the $p_{x}$ and $p_{y}$ characters vary in an opposite
manner to $p_{z}$, such that $p_{x}+p_{y}\simeq0.9-p_{z},$ which means that
$p_{x}+p_{y}$ is constant along the line $\theta=\pi/2.$ In the [110] hard
direction where $\theta=\pi/2$ and $\varphi=\pi/4$, the wavefunction should
extend mainly in the (110) plane, which can then be seen by approximately
equal characters of $p_{x},$ $p_{y}$ and $p_{z}$.

Because of the spin-orbit interaction, the levels do not have definite spin.
The highest occupied level acquires a minority-spin component that varies
between 6-8\% while the acceptor has a much smaller minority spin component
between 0.2-0.3\%. The shallower the impurity level, the more minority-spin
character is acquired; the second highest occupied level has larger minority
spin character in the range 12-16\%.

\subsection{Single Mn in the (110) GaAs surface layer}

\label{results_surface}

\begin{figure}[ptb]
\resizebox{7.5cm}{!}{\includegraphics{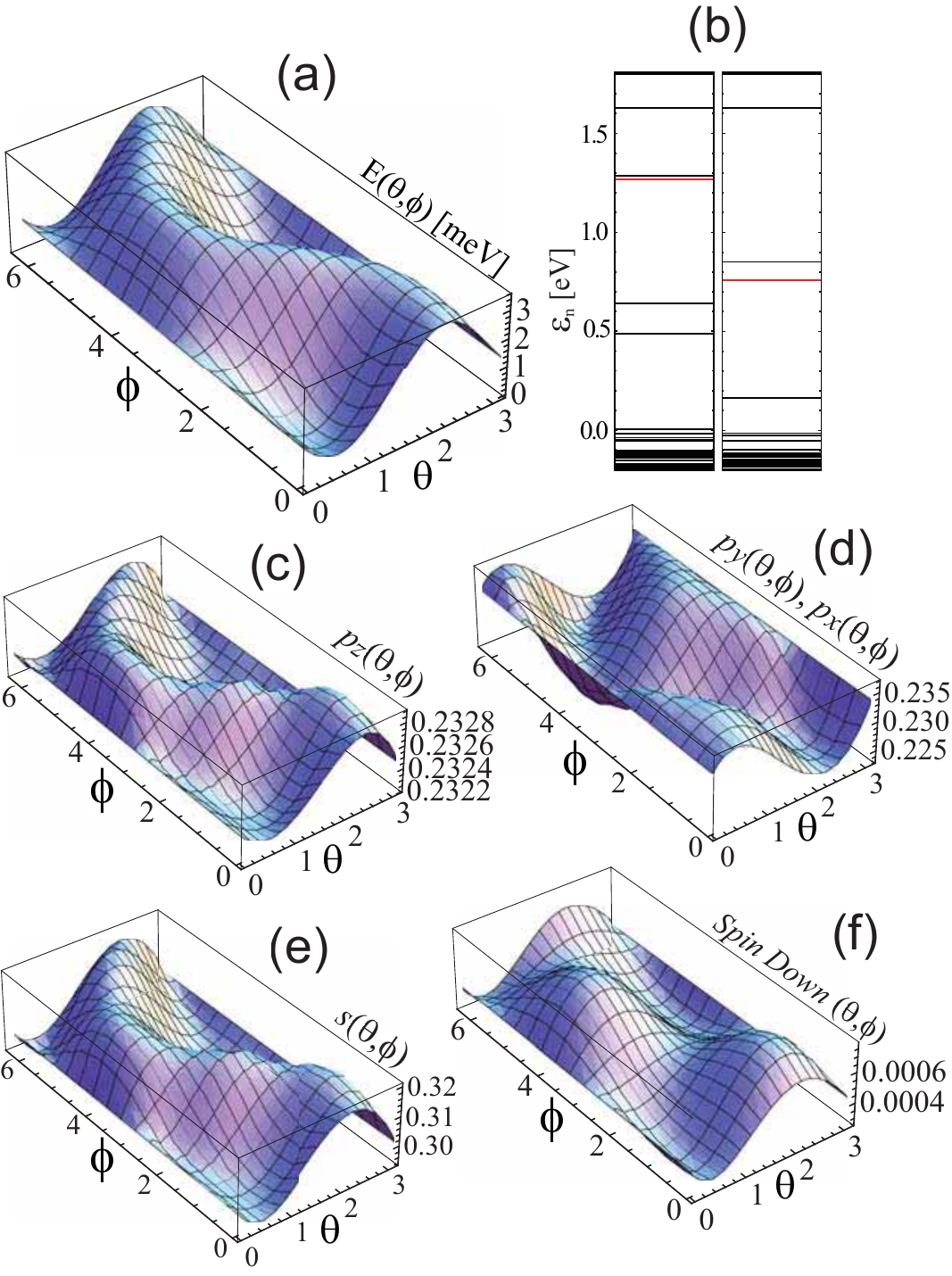}}\caption{(Color online) The
magnetic anisotropy energy of a single Mn on the (110) surface layer. Panel
(a) shows the magnetic anisotropy energy with bistable easy directions at an approximate 45
degree angle to the (110) surface in the [111] direction, separated by a
barrier of 3.5 meV. In (b), on the left, the eigenlevel spectrum in the easy
direction is shown, with the highest occupied level indicated by a red line. The
surface acceptor level is now deep in the gap at 1.27 eV. Reducing the
off-site Coulomb correction to 1.57 eV, causes the acceptor level to lie less
deep at the experimental position of the resonance at 0.85 eV (right hand
spectrum in (b)). The anisotropy correlates with the variation in the acceptor
orbital and spin character, shown in panels (c)-(f). }
\label{surfacemae}
\end{figure}\begin{figure}[ptb]
\resizebox{8cm}{!}{\includegraphics{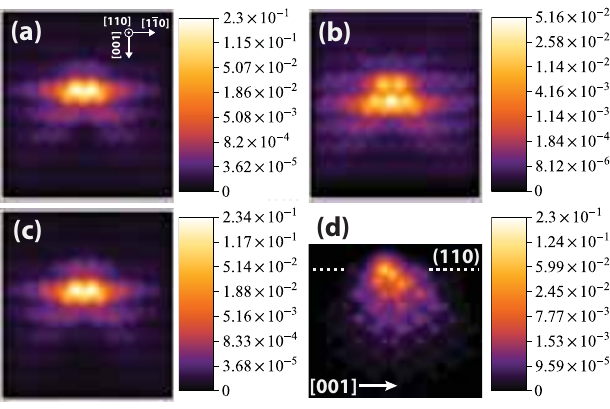}}\caption{(Color online) The
LDOS for a Mn at the (110) surface. (a) shows the LDOS at the surface and (b)
in the subsurface when the Mn spin is pointing in the easy direction. The
acceptor LDOS is much more localized than the bulk acceptor LDOS, with a total
of 63\% spectral weight on the Mn and its 3 neighboring As. (c) shows the
effect of reducing the off-site Coulomb correction in order to reproduce the
surface acceptor level at the experimentally observed position of 850 meV. 
Although this change shifts energies appreciably, it produces 
only very small changes in the spatial pattern of the LDOS. The star-like
shape found in experiment\cite{koenraad_prb08,yazdani_nat06} can be
distinguished in (a) and (c). (d) shows a maximal intensity projection (see main text for definition)
of the whole cluster along the line of sight (the [\={1}10] direction) and reveals
the symmetry responsible for the minima along [111]. }
\label{surfaceldos}
\end{figure}We now turn to the case of a single Mn in the $\left(  110\right)$ 
surface layer which has three As nearest neighbors, two located on the
surface -- see Fig.~\ref{clusterfig}.
In the calculations we use the parameters in the Hamiltonian that were
obtained by fixing the correct acceptor binding energy in bulk.
The magnetic anisotropy landscape on the unit sphere is shown in
Fig.~\ref{surfacemae} (a). At the surface, the anisotropy energy range is
$E_{\mathrm{anis}}=3.5$ meV and the landscape has bistable minima and an easy
axis at an approximate 45$
\operatorname{{{}^\circ}}
$ angle to the $\left(  110\right)  $ surface, corresponding to the [111] direction.
Fig.~\ref{surfacemae} (b) shows a portion of the eigenlevel spectrum.
We find that the acceptor level of a surface Mn is very deep in the gap. The
surface Mn acceptor level is not at all similar to its bulk counterpart, in
sharp contrast with what has previously been assumed. The highest occupied
level is also deep in the gap 12-19 meV below the acceptor level. The loss of
coordination at the surface is primarily responsible for the much deeper
state. A deep acceptor is also observed in experiment\cite{yazdani_nat06},
where the $dI/dV$ curve reveals a broad resonance at 850 meV above the valence
band edge. Our results imply that the acceptor at the surface can be
categorized as an intrinsically deep state, although the exact position
depends on the degree of $p$-$d$ hybridization at the surface.

On the basis of generic considerations which recognize the reduced symmetry at
the surface, one might have expected that the anisotropy would be highest at
the surface. Due to the nature of the Mn interaction with neighboring As, we
find that this is not the case. A high anisotropy requires a hole wavefunction
that is more spread out in the lattice causing high variations in the orbital
and spin character of the hole state. Fig.~\ref{surfacemae} (c)-(d) reveals
variations in $p_{x}$ and $p_{y}$ orbital character of 1\% and just a fraction
of a percent in $p_{z}$ character. This is in turn connected with the LDOS of
the hole that is more localized than bulk. Fig.~\ref{surfaceldos} (a) shows
the acceptor LDOS of the surface layer. The maximum spectral weight of 23.0\%
is not located on the Mn (which has 12.0\% spectral weight), but rather on
its 2 surface As nearest neighbors. The third As nearest neighbor in the
sublayer has a much smaller spectral weight of 5.2\%, which means that the
core region of the Mn and its 3 neighbors contains a total of 63\% spectral
weight. The Ga atoms have a smaller maximum spectral weight of 6.1\% for the
top layer and 2.0\% in the sublayer (see Fig.~\ref{surfaceldos} (b)). Although
the LDOS is highly localized, a similarity with the star-shaped symmetric
images observed in the STM topographs\cite{koenraad_prb08,yazdani_nat06} can
be seen. The presence of Zn dopants in experimental samples might increase the
coupling to conduction states, causing a more extended surface wavefunction.
The surface and subsurface LDOS is similar in easy and hard magnetization
directions; the changes in energy with magnetization direction are due to the
changes in $p_{x}$, $p_{y}$ and spin character on the same atoms.

The position of the acceptor state is sensitive to the off-site Coulomb
correction. Setting this parameter to reproduce the experimental position
gives an offsite Coulomb correction $V_{\mathrm{off}}=1.57$ eV. The
qualitative properties of the state does not change but the magnitude of the
anisotropy drops to 0.87 meV as we move away from the conduction band and the
gap between the highest occupied and the acceptor level increases.
Fig.~8 (c) reveals that this state has a very similar acceptor
surface LDOS.  
Our calculations indicate that the acceptor state for a single Mn on the (110)
surface is a deep, highly localized state with a relatively low anisotropy.
Precisely how deep and how localized this acceptor level is, is dependent on
model parameters that we must choose phenomenologically. It is possible that
the experimental surface Mn acceptor level binding energy quoted above could
be inaccurate because of band-bending effects which could be present when
performing STM on a semiconducting surface. Experimental studies of surface Mn
in more heavily doped (Ga,Mn)As samples, in which band-bending effects are
weaker could help settle this question.

We can understand the occurrence of the bistable minima approximately along [111] from
Fig.~\ref{surfaceldos} (d), which shows a maximal intensity projection from a
side view of the Mn, along the [\={1}10] direction. The presence of the
surface, causes the wavefunction to spread out in the (111) plane down into
the lattice. The star-like protrusions across rows observed in experiment are
weak surface echoes of this spread. 
The maximal intensity projections are generated
as follows. Instead of placing two dimensional Gaussians at the atomic sites (like  in Fig.~\ref{surfaceldos} (a)-(c)), we place a Gaussian sphere at each atomic site $\vec{x}_{i}$: $l_{i}e^{(\vec{x_i}-\vec{r})^2/\Gamma^2}$. Here, the magnitude of the Gaussian $l_{i}$ is equal to the LDOS value of atom $i$ (as obtained from Eq.~\ref{projatom}), and we choose the Gaussian smearing $\Gamma$ such that the full width at half maximum is equal to half the nearest neighbor distance. This generates a three dimensional LDOS view of the entire supercell cluster, with one color value corresponding to the LDOS for each point in space.
The maximal intensity projection takes the maximum LDOS value along the viewers line of sight (in this case the [\={1}10]) and projects it onto the viewing plane, i.e. the plane which is perpendicular to the viewers line of sight. In this way, we obtain an image which is not directly related to the STM images, but provides useful information on the LDOS below the surface. 

\subsection{Single Mn in subsurfaces layers of GaAs(110)}

\label{surface_to_bulk}

This subsection is devoted to the study of what happens when the Mn is
successively moved down from the surface layer, toward the cluster center
layer which best approximates the bulk. \begin{figure}[ptb]
\resizebox{8.2cm}{!}{\includegraphics{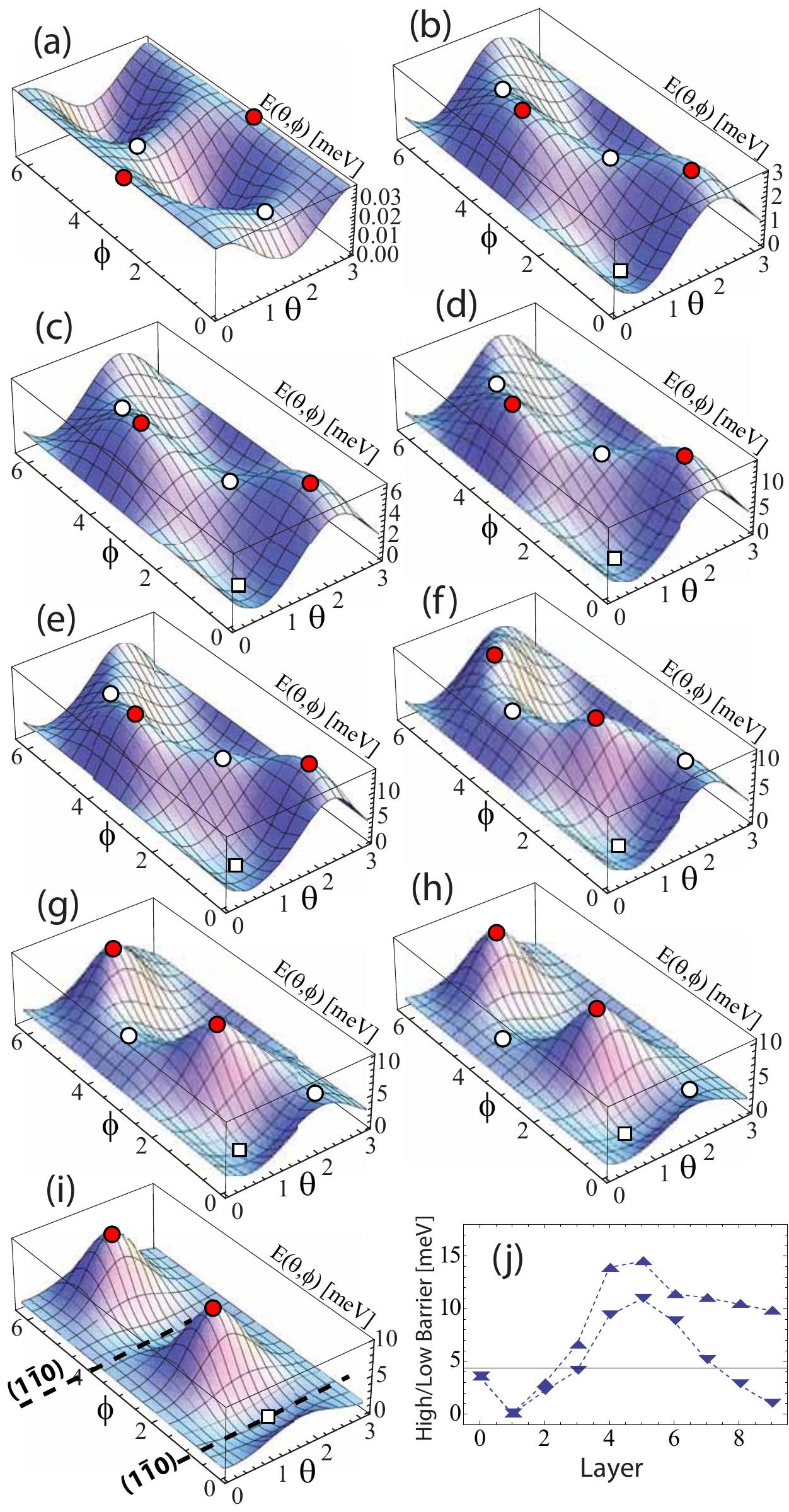}}
\caption{(Color online) Magnetic anisotropy over the unit sphere as a function of depth.
(a)-(i) correspond to the anisotropy landscapes of sublayers 1-9. Sublayer 1
has extremely low anisotropy, with easy directions along the [1\={1}0].
In sublayer 2 (panel (b)) a surface-like landscape reappears, but
with a high and a low blocking barrier. The low barriers are marked with an open
circle, the high barrier (hard) directions with a filled circle and the minimum energy easy directions
with a square. The barriers both grow with depth
until they exchange positions in sublayer 6. The anisotropy energy reaches a
maximum of 15 meV in sublayer 5 in (e). The difference between the high and
the low barrier then increases with depth until a quasi-easy (1\={1}0) plane
forms (marked with dashed line). (j) tracks the high and the low barrier as a function of the Mn depth,
where the horizontal line indicates the bulk blocking barrier.}
\label{subanis}
\end{figure}Fig.~\ref{subanis} shows how the anisotropy energy varies over the
unit sphere with increasing depth, with panels (a)-(i) corresponding to
sublayers 1-9. The subsurface Mn landscape in Fig.~\ref{subanis} (a) reveals
an extremely low anisotropy energy of 30$\mu$eV. As in the surface layer case,
the small anisotropy can be traced to a very low variation in orbital
character associated with a highly localized hole wavefunction. In the second
sublayer (Fig.~\ref{subanis} (b)), the Mn produces an anisotropy landscape
similar to that of the surface, but with high and low blocking barriers of
2.2-2.9 meV. In the successive layers, the high and the low barriers both
increase and become larger than bulk, reaching a maximum of 14.5-10.9 meV in
sublayer 5 (Fig.~\ref{subanis} (e)). In sublayer 6 (Fig.~\ref{subanis} (f))
the high and the low barrier have interchanged positions.
The easy axis remains approximately along the [111] all the way down to sublayer 6
where it begins to shift towards the surface normal. The low barrier decreases
towards the center and at the deepest layer of the slab the low barrier
becomes so low that an approximate easy (1\={1}0) plane forms (see
Fig.~\ref{subanis} (i) at $\phi=\pi/4,5\pi/4$), opening up reversal paths
connecting the two bistable minima. 
In the deepest sublayers the Mn spin 
can explore the plane with $\phi=\pi/4,5\pi/4$ and any $\theta$
with a very low energy cost. This plane corresponds to (1\={1}0), which
can then be classified as a quasi-easy plane. 
At the deepest levels, we still see traces
of the now very shallow surface minima. 

\begin{figure}[ptb]
\resizebox{5.7cm}{!}{\includegraphics{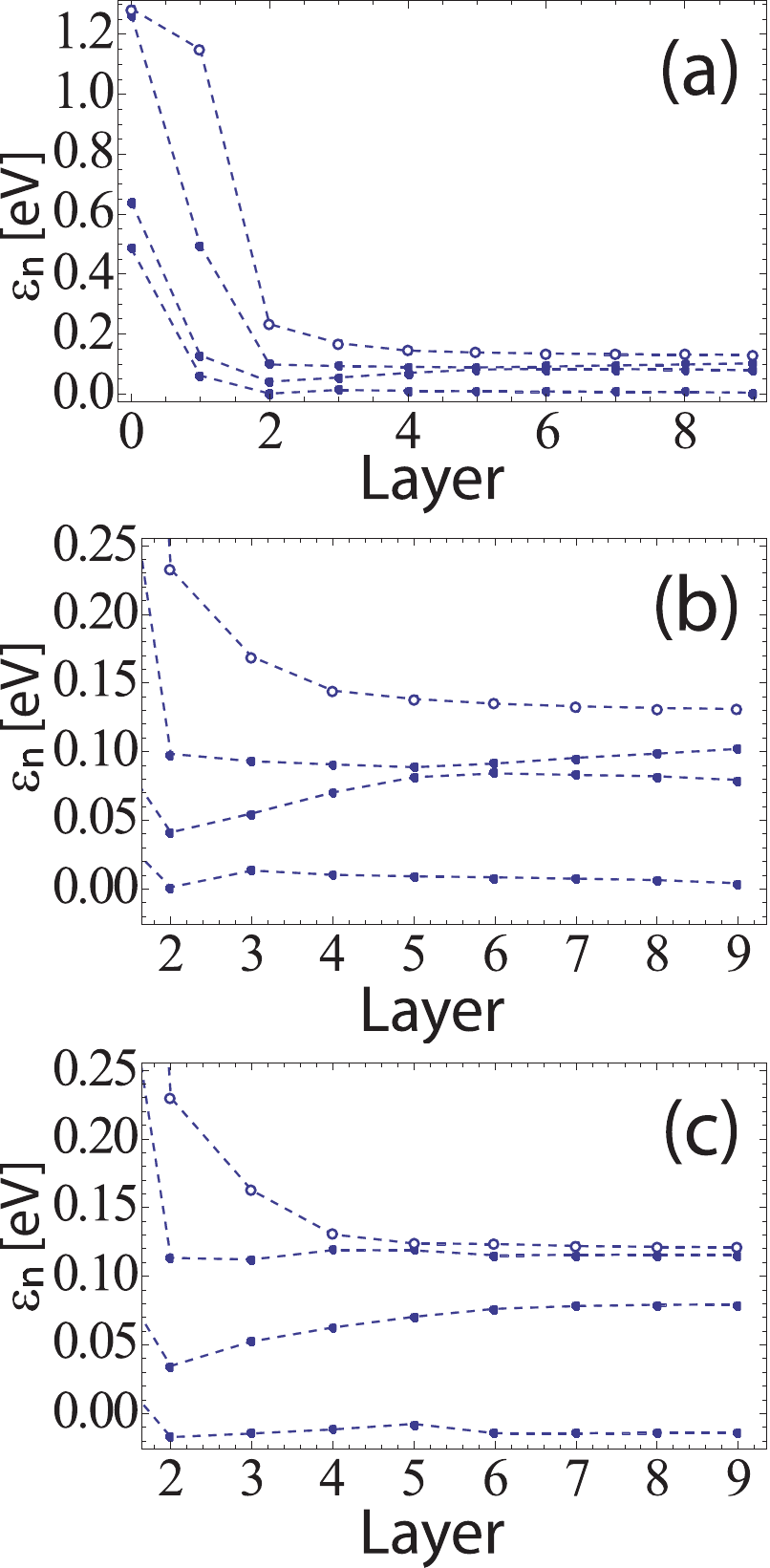}}
\caption{Energies of the three highest occupied levels (filled circles) and the acceptor level (empty circles) as a function of Mn depth, starting from a Mn on the surface (layer 0), when the Mn spin is pointing in the easy direction (marked by squares in Fig.~\ref{subanis}). The acceptor abruptly drops down towards the valence band in sublayer 2 and then flattens out, converging at 125 meV above the first valence band level, corresponding to a slightly deeper acceptor than in bulk. (b-c) Energies of these four levels as a function of sublayer index, starting from sublayer 2; (b) is for the easy direction and (c) for the hard direction, respectively. The hard direction is the magnetization direction corresponding to the high barriers (marked by filled circles in Fig.~\ref{subanis}).}
\label{sublevels}
\end{figure}

Fig.~\ref{sublevels} (a) shows the evolution of the three highest occupied levels and the acceptor level as a function of Mn depth, when the Mn spin is pointing in the magnetic easy direction (marked by squares in Fig.~\ref{subanis}). The acceptor abruptly drops down towards the valence band in sublayer 2 and then flattens out, converging at 125 meV above the first valence band level, corresponding to a slightly deeper acceptor than in bulk. In Fig.~\ref{sublevels} close-ups of the same energy levels is shown in the easy (b) and hard (c) direction, respectively. The hard direction is given by the magnetization direction corresponding to the high barriers marked by a filled circle in Fig.~\ref{subanis}.

The difference between the acceptor
levels in the hard and easy direction is the spin-orbit induced total
anisotropy energy. The maximal anisotropy in sublayer 5 corresponds to a very large
variation in the gap between the highest occupied level and the acceptor of
5-50 meV between the hard and the easy direction. The two quasi-degenerate
states are split by the spin-orbit shift.
The easy direction is the direction in the which the gap caused by this shift,
is maximized, leading to a decrease of the total energy of the system. A plot
of the high and the low barrier as a function of sublayer depth is shown in
Fig.~\ref{subanis} (j). From this figure we see the magnetic anisotropy energy
is maximally enhanced in sublayer 5. This is due to the presence of the
surface and has to do with the way the acceptor wavefunction extends around
the Mn, as we move down through the layers. 
\begin{figure}[ptb]
\resizebox{7.0cm}{!}{\includegraphics{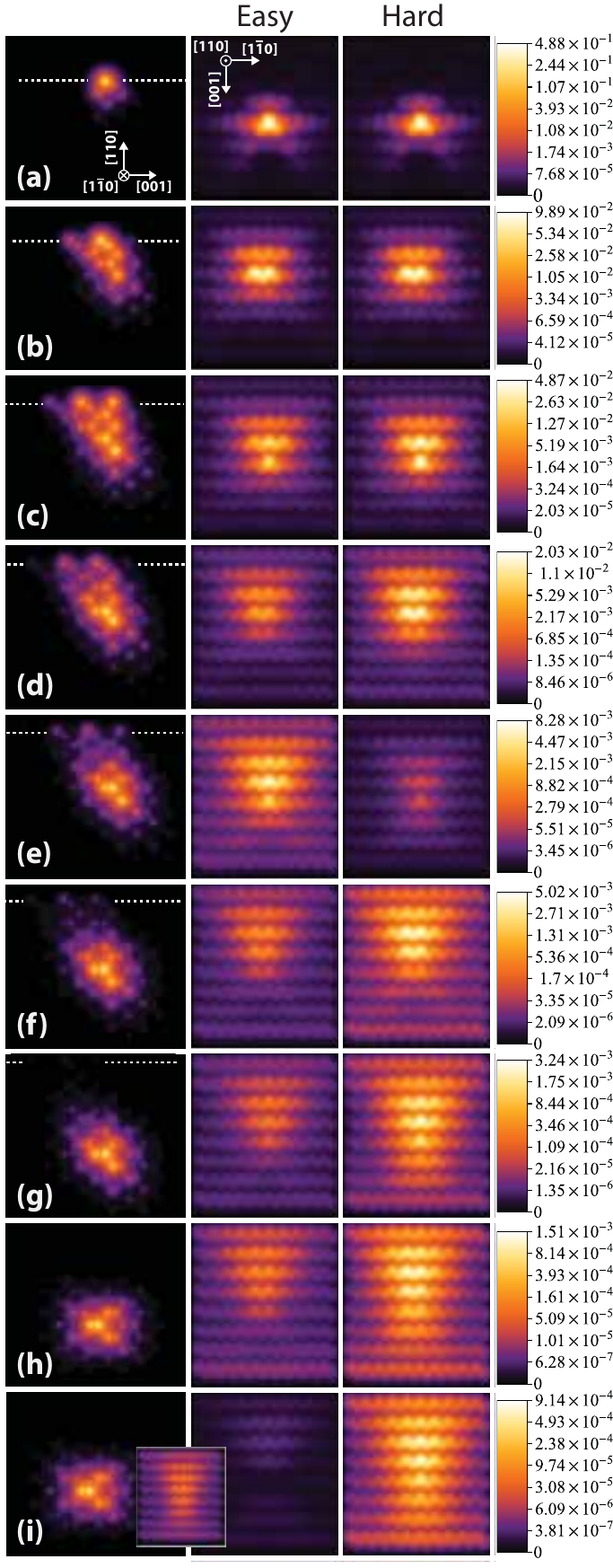}}\caption{(Color online) The (110)
surface LDOS as a function of the depth of Mn sublayer. (a)-(i) correspond to
sublayers 1-9. Maximal intensity projections along [1\={1}0] in the left
column show how the LDOS in the ground state solution eventually detaches from
the surface, suppressing the extension along the [110] direction. The maximal
intensity projections show the relative distribution of the spectral weight in
the cluster (thermometers apply to center and right column).
The two right columns show the surface LDOS in the easy and hard
direction (see Fig.~\ref{subanis}). The hard direction generally has a higher
LDOS maximum and its pattern shows more spectral weight on the [001] side with increasing depth. 
The inset in (i)
shows the surface LDOS when the Mn spin is pointing in the [001] direction,
which is very close in energy to the easy direction. }
\label{subcuts}
\end{figure}

Fig.~\ref{subcuts} shows how the LDOS in the (110) surface plane evolves as a
function of the sublayer depth in the hard and easy direction respectively.
This figure therefore relates directly to the window on acceptor level
properties opened by STM experiments. 
The left column of Fig.~\ref{subcuts} shows maximal intensity projections (see definition above)
for the easy direction solution. This series of images gives a qualitative
idea of where the main spectral weight of the acceptor wavefunction is located in
relation to the surface (indicated by dashed line). We see that when the Mn impurity is close to the
surface, the buckling has a large effect on its spatial extension.
As the acceptor wavefunction detaches from the surface with increasing depth, the extension towards
the surface becomes reduced and it begins to extend along [1\={1}0] (not shown in this
sequence of images as the extension along [1\={1}0] is parallel to the line of sight).

The center and the right column of
Fig.~\ref{subcuts} shows the (110) surface LDOS when the Mn spin is pointing
in the easy and hard direction, as depicted in Fig.~\ref{subanis}.
To begin with, we examine these images from a qualitative point of view.
Starting from the case in which the Mn is inserted in the second sublayer or
deeper (Fig.~\ref{subcuts} (b)-(e)), the LDOS on the (110) surface displays a
characteristic triangular shape with one vertex pointing down in the [001]
direction. As the Mn is inserted into yet deeper layers, the triangular shape
evolves into shape resembling a butterfly or bow-tie, with stronger
\textit{upper} wing. These features have been observed
experimentally\cite{koenraad_prb08, jancu_prl08} and found in agreement with
TB calculations\cite{jancu_prl08} similar to ours. It is interesting to notice
that for a Mn in bulk, calculated cross-section LDOS on (110) plane located
$n$ atomic layers from the Mn shows a similar butterfly shape, but with a
stronger \textit{lower} wing.\cite{jancu_prl08} Our calculations for bulk Mn
shown in Fig.~\ref{bulkldos} displays a similar tendency -- see in particular
Fig.~\ref{bulkldos} (d)-(e). This symmetry reversal of the bow-tie shape with
respect to the [001] direction for a Mn close to the surface has been ascribed
to the intrinsic strain associated with the buckling
relaxation.\cite{jancu_prl08}
The change of the Mn-induced LDOS from a triangular to a bow-tie shape as the
impurity is inserted into deeper layers below the (110) surface occurs after
sublayer 5, where the magnetic anisotropy landscape starts to develop a
quasi-easy plane. The Mn spin is now free to explore the easy-plane (1\={1}0),
with in more spectral weight in the [1\={1}0] direction in the Mn layer below the surface.
We now proceed with a more
quantitative evaluation of Fig.~\ref{subcuts}. The LDOS of the first sublayer
(Fig.~\ref{subcuts} (a)) is highly localized with a maximum spectral weight on
the surface As of 50\%. The Mn, one layer below, has only 8\% spectral weight.
The change in the LDOS with the Mn spin direction is minute and the
resulting very low variations in orbital and spin character yield a low
anisotropy energy. It is clear that the rising anisotropy in sublayer 2 and
below is associated with a generally more spread out acceptor wavefunction. In
sublayer 2 (Fig.~\ref{subcuts} (b)), the hole is now much more spread out than
at the surface and subsurface, and it shows that a large part of the extended
spectral weight lodges at the surface. In sublayer 1 and 2 there is little
change between the hard and easy direction, and in sublayer 3 and 4 the
maximum spectral weight going from the hard to the easy direction, decreases
by 14\% and 33\%. In sublayer 5 (Fig.~\ref{subcuts} (e)), where the maximum
anisotropy energy is reached, the easy direction surface LDOS instead becomes
more pronounced and the maximum LDOS decreases by 60\% from the easy to the
hard direction. There is clearly something special about sublayer 5, where the
surface has a high impact on the anisotropic extensions of the acceptor
wavefunction. In addition to the large change in magnitude between the hard
and easy directions when the Mn is in this particular layer, the surface LDOS
begins to show a qualitative change.
In sublayer 6 (Fig.~\ref{subcuts} (f)), at which the high and low barriers are
first interchanged, the wavefunction again shows a stronger surface LDOS in
the hard direction. As we move further below the surface, the LDOS maximum
decreases by around 40\% as the Mn spin reorients from the hard to the easy
direction. At the deepest level in sublayer 9 (Fig.~\ref{subcuts} (i)) the
quasi-easy (1\={1}0) plane has formed, and the surface LDOS maximum abruptly
drops 82\% between the hard and easy direction. The general trend in acceptor
wavefunction character as layer depth is increased is that the surface LDOS is
decreases and extends more along the [1\={1}0] direction, running below the surface.

Although the surface LDOS is biased on the [00\={1}] side of the Mn, the
patterns are consistent with the fact that the wavefunction tends to extend
along directions perpendicular to the Mn spin. In the hard directions (where
the Mn spin is in the [11\={1}] or the [1\={1}0]), we therefore consistently
see more spectral weight on the [001] side of the Mn relative the easy direction LDOS,
and a more pronounced bow-tie like pattern.
For the deeper levels where the low barrier has dropped significantly, thermal
and quantum fluctuations, as well as Mn-Mn interactions in the sample can
cause the Mn spin to fluctuate in the (1\={1}0) plane, such that the bow-tie
shape becomes more pronounced. The small inset in Fig.~\ref{subcuts} (i) shows
the LDOS at the surface when the Mn spin is in the [001] direction, which has the familiar
slightly asymmetric bow-tie shape. This solution is very close in energy to
the easy direction, only 1 meV higher. For deep impurities, the wavefunction
is essentially dominated by the subsurface extension along the [1\={1}0], such
that the Mn spin can move across the low barrier in the (1\={1}0) plane.
Comparing with the fully periodic system, where the wavefunction extends
equally along the [1\={1}0] and the [110] in the ground state, we see that the
effect of the surface on the deep impurities is to reduce the extension along
the surface normal, such that reversal paths open up in the single barrier for
bulk (see Fig.~\ref{bulkmae}). It is also noteworthy that in sublayer 5, the
low and the high barrier has not yet interchanged, but are comparable in
energy. The LDOS at the surface when the Mn spin is pointing in the direction
of the lower barrier is larger than the LDOS in the easy direction, then
following the same qualitative pattern as all the other depths. This
indicates that the wavefunction is very sensitive to the level dynamics as the gap between the highest occupied and the acceptor level closes in. The odd behavior of sublayer 5 is associated with a quasi-degeneracy between the highest occupied and the acceptor level, as indicated by the large variations 5-50 meV of the gap between them (see Fig.~\ref{sublevels}). This sensitive situation, causes a large total amount of spectral weight to be shifted on and off the surface between the hard and easy direction, such that
large variations in orbital and spin character occur, yielding the high
anisotropy.

\begin{figure}[ptb]
\resizebox{6.3cm}{!}{\includegraphics{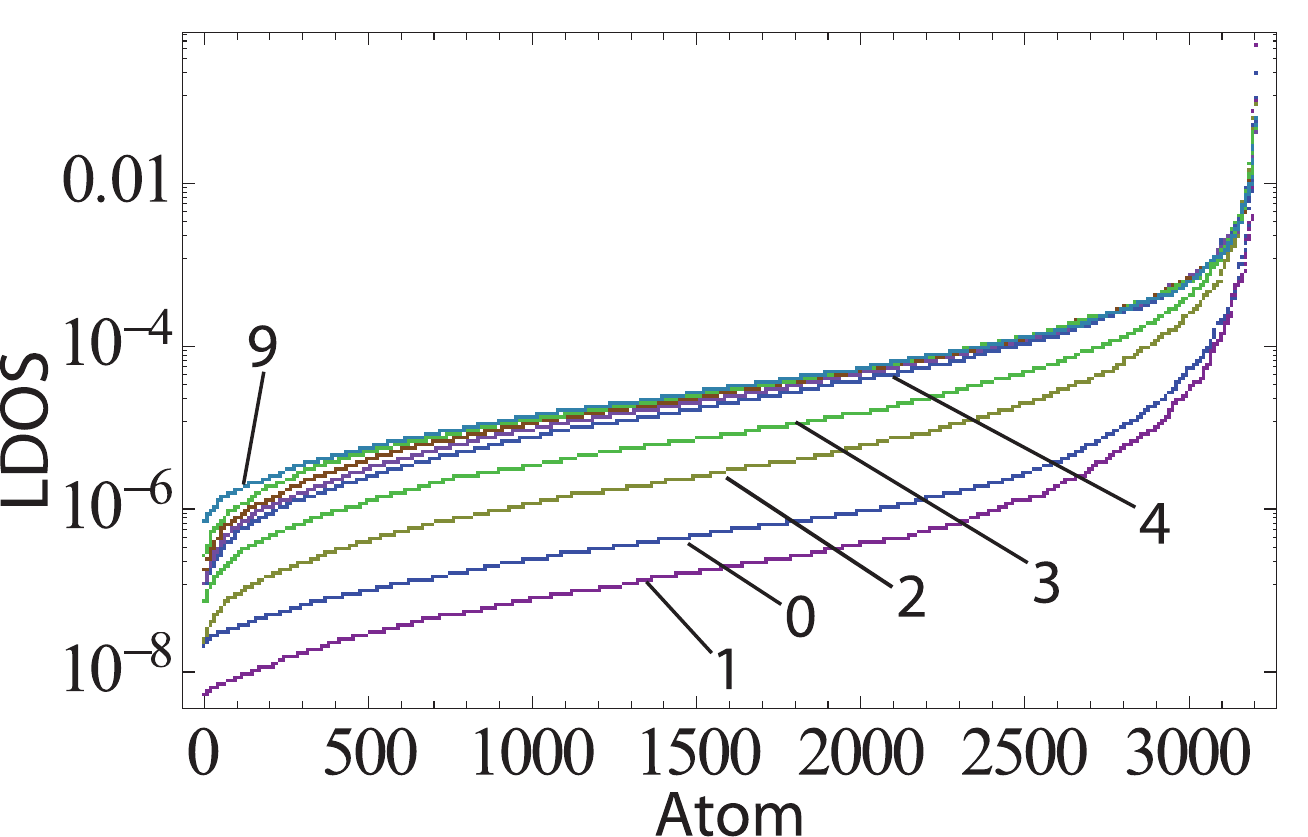}}\caption{(Color online) The
LDOS of every atom in the cluster for the acceptor level for different Mn depths (marked 0-4,9). The
graph shows the 3200 values of LDOS (one for each atom) sorted in size  
and plotted on a logarithmic
scale. The surface (marked 0) and the subsurface (1) both show a similar,
highly localized distribution with a large population that has very low spectral weight
in the range $10^{-8}-10^{-7}$. 
As the Mn is placed in deeper layers, the
acceptor LDOS becomes more extended with a majority population of atoms in the range $10^{-4}-10^{-6}$.}
\label{ldosdistributions}
\end{figure}

In Fig.~\ref{ldosdistributions} the acceptor level LDOS values on all 3200 atoms
are sorted in size and then plotted on a logarithmic scale. 
Note that what is shown is not a continuous curve illustrating a parameter dependence but 3200 closely spaced points,
one for each atom. The reason that the values have been sorted in size is that we are only interested in the possible
magnitudes of the atomic LDOS here. There is no spatial information in this plot, but it shows how the LDOS magnitides are spread
over the population of atoms. The sum of the LDOS values is normalized to unity.
This plot demonstrates that an increased depth of the Mn is associated
with a delocalization of the acceptor wavefunction, which manifests itself as a large population of atoms with increased spectral weight. 
As the Mn depth increases a large population of atoms emerges with approximately 2-3 orders of magnitude larger LDOS relative the highly localized surface set. 
The surface (marked 0) and the first sublayer (marked 1) exhibit similar distributions, with a localized signature and a larger spectral weight in the very narrow high end limit on the far right. For the surface and subsurface, the bulk of the spectral weight is located in close vicinity to the Mn, which results in a large population of atoms with a much lower LDOS in the range $10^{-8}-10^{-7}$.
For sublayer 3 and deeper, the majority of atoms show increased weight in the approximate range $10^{-6}-10^{-4}$. This large population represents atoms farther away from the Mn core region, which means
that the acceptor wavefunction is becoming much more extended away from
the surface.

\begin{figure}[ptb]
\resizebox{6.3cm}{!}{\includegraphics{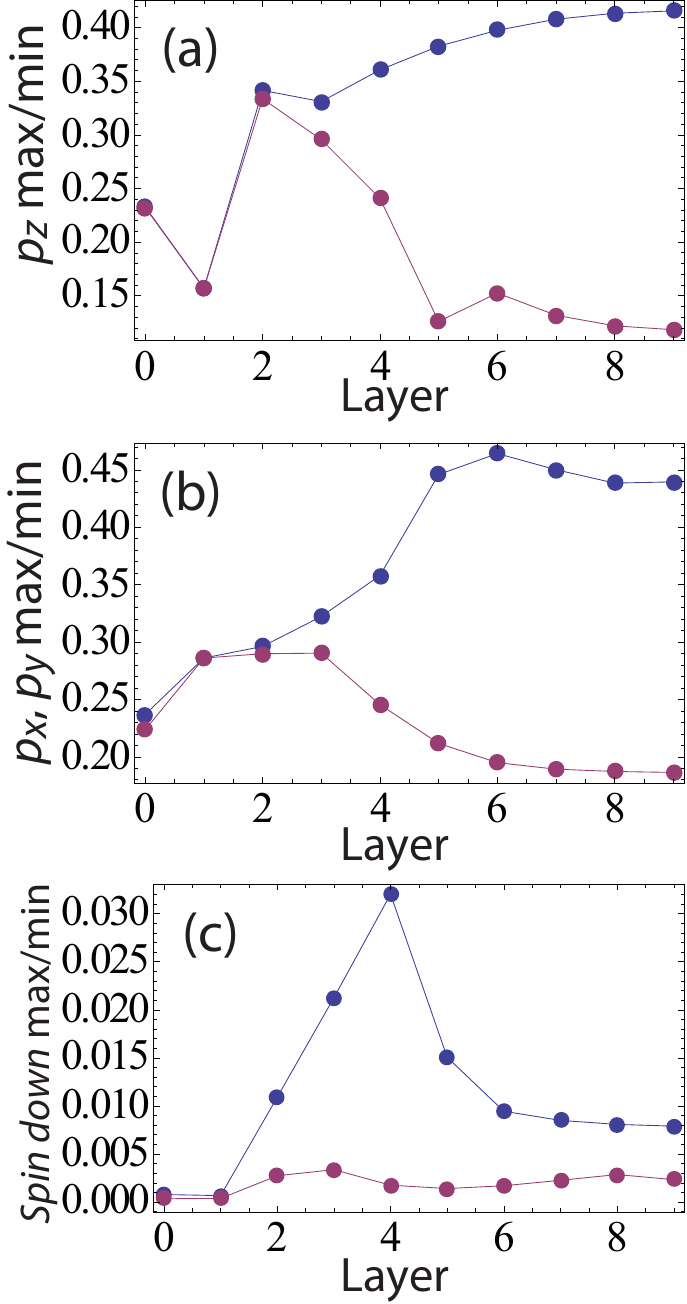}}\caption{The
evolution of the maximum and minimum orbital and spin character of the
acceptor level on the unit sphere as a function of Mn depth. Orbital and spin
character fluctuations are small in the highly localized surface and
subsurface wavefunctions. They then increase to very large variations in
sublayer 5, after which they level out. The spin down character follows a
similar pattern, but drops sharply after sublayer 4. }
\label{orbstats}
\end{figure}The spatial spread of the acceptor wavefunction is connected with
the variations in orbital and spin character. Fig.~\ref{orbstats} shows the
maximum and minimum orbital and spin down character on the unit sphere of the
Mn magnetic moment directions as a function of Mn depth for the acceptor
level. Layer 0 (the surface) and layer 1 (the subsurface) show little
variation. The difference between maximum and minimum orbital character
rapidly increases from sublayer 3 and reach high values in sublayer 5 and 6,
where the anisotropy energy is also large. As we move further down, the
orbital $p$-characters level out towards a 25\% maximum and minimum
difference. 
As can be seen in Fig.~\ref{sublevels} the acceptor level comes very close to the highest occupied level in sublayer 5, leading to a quasi-degeneracy between these two levels in the hard direction. This leads to a large anisotropy energy, as the quasi-degeneracy is lifted and the total energy lowered, when the Mn spin is pointing in the easy direction. Associated with this quasi-degeneracy, is a mixing of the two levels and a change in the nature of the acceptor wavefunction. This change manifests itself as large fluctuations in orbital and spin character of the wavefunction in sublayer 5, as can be seen in Fig.~\ref{orbstats}. The quasi-degeneracy between the acceptor and highest occupied level in sublayer 6, is also related to the change in orbital and spin character responsible for the interchange of the high and low barriers as we go from sublayer 5 to 6 (see Fig.~\ref{subanis} (e-f)).
The spin-down character variation (Fig.~\ref{orbstats} (c)) increases steeply in sublayer 2 and reaches a maximum in sublayer 4, after which the difference drops quickly towards 5\%. The anisotropy is also quite high in sublayer 4, but the orbital character variations are not as large as in sublayers 5 and 6. However, looking at the variations in spin-down character, we see that there is a sharp maximum on sublayer 4. This points to the crucial impact of small spin character fluctuations of the acceptor level on the magnetic anisotropy energy.

Fig.~\ref{ldosdistributions} and \ref{orbstats} show that the acceptor
wavefunction delocalizes with depth, leading to larger variations in orbital
and spin character as the spread increases. Because of the surface, the
variations in spin and orbital character are related to the geometry of the
system that is responsible for redistribution of the LDOS between the hard and
easy directions. The total redistribution is very large for the maximal
anisotropy layer, essentially depleting the surface LDOS in the hard direction
to the advantage of the LDOS in the easy direction.

\section{Summary and Conclusions}

\label{conclusions}

We have undertaken a study of the properties of a single Mn embedded in a
large 3200 atom GaAs matrix. 
Our model is based on a kinetic-exchange
tight-binding Hamiltonian that accounts for the polarization of the As $p$-electrons
nearest neighbors of a Mn spin via an off-site exchange term, and includes the
local atomic spin-orbit interaction as well as the Coulomb field from the Mn
ion. 
The relaxation of the (110) surface was taken into account 
since it plays an essential role\cite{yazdani_nat06,koenraad_prb08} in the (110)
surface electronic structure.

In agreement with previous calculations\cite{tangflatte_prb05} we find that
the acceptor wavefunction is generally highly dependent on the Mn spin
direction. An exception occurs for the cases of surface and subsurface layer Mn spin locations,
for which we find a highly localized acceptor wavefunction that does not spatially
redistribute as the Mn spin changes direction. We find, in particular, that a single Mn
in the (110) surface produces a highly localized acceptor-level wavefunction 
and that the level lodges
deep in the gap due to the loss of coordination at the surface. A deep
acceptor level at the surface is also found in experiment\cite{yazdani_nat06},
although uncertainly about band-bending effects has been responsible
for some confusion as to its energetic position.
The anisotropy of the surface state is relatively low, as expected for a
deep-gap state. The actual depth of this state in our model depends
sensitively on a purely phenomenological parameter in our calculation, 
the off-site Coulomb correction, which needs
to be reduced relative to its bulk value 
in order to reproduce the experimentally observed position in
the gap.

The acceptor level spatial structure is probed experimentally via the LDOS contributions to the surface layer.  
As the Mn is placed in deeper subsurface layers, the surface LDOS 
first displays a triangular pattern, which then evolves
into an asymmetric butterfly or bow-tie shape when the Mn is located still further
from the surface. This finding is in qualitative agreement with
experiment.\cite{koenraad_prb08,jancu_prl08} We find in addition that the 
anisotropy energy grows with depth, as the acceptor wavefunction becomes more
extended. The magnetic anisotropy easy
direction is strongly affected by the surface, and we find that its presence
tends to favor [111] as an approximate easy direction. The observed easy axis is
associated with a tilt of the wavefunction due to the up and down shift of the
surface As and Ga atoms. This symmetry is responsible for the triangular
pattern on one side of the Mn and persists to deep layers, after which the Mn
spin becomes more free to move across the surface normal. 
By comparing a three-dimensionally 
periodic calculation with the corresponding wide surface slab calculations, 
we see that for the deeper Mn, the effect of the surface is to 
open up additional reversal paths in the bulk Mn impurity's single-barrier magnetic anisotropy landscape.
This is because at the deepest layers, the ground state solution yields an acceptor 
wavefunction that is suppressed along the surface normal [110], 
relative to the fully periodic system.  The deep layer wavefunction
extends instead most strongly along the [1\={1}0] direction, such that a quasi-easy [1\={1}0] plane is formed. 
A strong magnetically and spatially
anisotropic state is found in sublayer 5, where a quasi-degeneracy between the acceptor and highest occupied level leads to a large anisotropy energy and a
large shift of the LDOS at the surface between hard and easy directions.

Our predictions can in principle be checked experimentally by manipulating the Mn
magnetic moment direction with an external magnetic field. As a function of
the Mn depth, effects could be visible as early as the third sublayer, where
our calculations indicate that the surface-layer LDOS for magnetization along the 
[11\={1}] hard direction should be larger than the surface layer 
LDOS when the magnetization is along the easy direction [111]. In sublayer 3
and 4 the maximum LDOS\ decreases by 14\% and 33\% from the hard to the easy
direction. The acceptor level becomes increasingly shallow with increasing Mn
layer depth and approaches the valence band in sublayer 5. We observe a strong
signature in sublayer 5.  At this point the surface-layer LDOS is instead larger for 
magnetization along the easy direction, with a 60\% decrease in maximum LDOS from the easy to the
hard direction. It should be noted however, that this situation depends very
sensitively on the local environment since this signature is associated with a
quasi-degeneracy between the highest occupied level and the acceptor level. For
instance, increased Mn doping can affect the layer index at which this change occurs. 
In sublayer 6, the high and the low barriers have
interchanged, with the high barrier now in the [1\={1}0] direction
parallel to the surface.  At this point the easy direction is still approximately the
[111], but it then moves towards [110] with further increased depth. In
addition to a large 40\% decrease in the maximum LDOS from the hard to the
easy direction, our calculations indicate that a more symmetric bow-tie shape
appears in the hard direction for layer 6 and deeper. 

The Mn spin-orientation can be influenced by thermal and quantum 
fluctuations as well as by external magnetic fields.  
We will address the influence of quantum fluctuations in a subsequent 
publication, and comment here only on the interplay between 
external fields and thermal fluctuations.  At temperatures higher than the 
anisotropy energy the Mn spin orientation will be randomized by interactions with 
its thermal bath. The measured surface-layer LDOS should then correspond to an 
average of the results obtained here for particular orientations.   
At low-temperatures thermal fluctuations become unimportant and the magnetization
orientation will depend on a competition between Zeeman coupling and 
magnetic anisotropy.  As we have explained, the surface-layer LDOS pattern
provides an indirect fingerprint of the magnetization orientation.  
Our experimental predictions for the depth dependence of magnetic anisotropy can be tested by 
identifying the Zeeman-coupling strength required to change the surface-layer LDOS.
For example we predict that for Mn in deeper layers 
there are high magnetic barriers of the order 10 meV, which
implies that large magnetic fields are required to reorient the Mn spin to the
hard directions.  These high barriers for a single Mn suggests that it might 
be possible to engineer few-atom substitutional Mn
impurity clusters near the GaAs surface which act like nanomagnets with 
attractively large magnetic blocking temperatures.

\section{Acknowledgments}

This work was supported by the National Science Foundation under
grant DMR-0606489, the Faculty of Natural Sciences at Kalmar University, the
Swedish Research Council under Grant No: 621-2004-4439, and by the Office of
Naval Research under grant N00014-02-1-0813. We would like to thank 
A.~Yazdani, P.~M.~Koenraad, J.~K.~Garleff, A.~P.~Wijnheijmer and C.~F.~Hirjibehedin for useful discussions.

\end{document}